\documentclass[preprintnumbers,amsmath,amssymb,preprint]{revtex4}

\usepackage{amsmath}
\usepackage{graphicx}
\usepackage{dcolumn}
\usepackage{bm}
\usepackage[usenames]{color}
\usepackage{multirow}

\begin{document}

\preprint{draft version}

\title{Robustness of Highly Entangled Multi-Qubit States Under Decoherence}

\author{A. Borras$^{1}$\footnote{toni.borras@uib.es},
A.P. Majtey$^{1}$\footnote{ana.majtey@uib.es},
A.R. Plastino$^{2,\,3,\,4}$\footnote{angel.plastino@up.ac.za},
M. Casas$^{1}$\footnote{montse.casas@uib.es}, and
A. Plastino$^{3}$\footnote{plastino@fisica.unlp.edu.ar}}
\affiliation{ $^1$Departament de F\'{\i}sica and IFISC, Universitat de
les Illes Balears, 07122 Palma de Mallorca, Spain \\
$^2$Instituto Carlos I de F\'{\i}sica Te\'orica y Computacional,
Universidad de Granada, Granada, Spain \\\\
$^3$National University La Plata UNLP-Conicet, C.C. 727, 1900 La
Plata, Argentina
\\\\$^4$Department of Physics, University of Pretoria -
0002 Pretoria, South Africa }

\date{\today}
\begin{abstract}
We investigate the decay of entanglement, due to decoherence, of
multi-qubit systems that are initially prepared in  highly (in some
cases maximally) entangled states. We assume that during the
decoherence processes each qubit of the system interacts with its
own, independent environment. We determine, for systems with a small
number of qubits and for various decoherence channels, the initial
states exhibiting the most robust entanglement. We also consider a
restricted version of this robustness-optimization problem that only
involves states equivalent, under local unitary transformations, to
the $|GHZ\rangle$ state.

\end{abstract}


\maketitle

\section{Introduction}

 Entanglement and decoherence are two intimately related
 phenomena that lie at the heart of quantum physics
 \cite{BZ06,NC00,S05}.  Entanglement is probably
 the most distinctive feature of quantum mechanics, its many
 manifestations being the focus of an intense and increasing
 research activity. From the fundamental point of view
 entanglement plays an important role, for instance, in explaining
 the origin of the classical macroscopic world from a quantum
 mechanical background \cite{S05}, and also in connection
 with the foundations of statistical mechanics \cite{GMM04,PSW06}.
 On the other hand, the creation and
 manipulation of multi-partite entangled states have important
 technological applications, such as quantum computation \cite{NC00}
 and quantum metrology \cite{GLM06}. The phenomenon of decoherence
 consists, basically, of a set of effects arising from the
 interaction (and concomitant entanglement-development)
 between quantum systems and their environments \cite{NC00,S05}.
 Almost every physical system is immersed in an environment and
 interacts with it in some way. The associated entanglement
 developed between the system and the environment leads to the
 suppression of typical quantum features of the system, such as
 the interference between different system's states. This
 constitutes the basis of the ``decoherence program" for
 explaining quantum-to-classical transition \cite{S05}.

  The (internal)  entanglement exhibited by a composite
  system undergoing decoherence tends to decrease as the
 process of decoherence takes place. This decay of entanglement
 has recently attracted the interest of many researchers
 \cite{YE04,QJ08,SK02,CMB04,DB04,HDB05,ACCAD08,GBB08,AlmeidaEtAl07,SallesEtAl08}
 because it constitutes one of the main obstacles for the development
 of quantum technologies based on the controlled  manipulation
 of entangled states \cite{NC00}. It has been shown that in some
 cases entanglement can vanish in finite times. This phenomenon
 is known as \emph{entanglement sudden death} \cite{YE04,QJ08}.
 Numerous works have been devoted to the study of the robustness of
 various multipartite entangled states under the influence of
 different decoherence models,  paying special attention to
 their scaling behavior with the size of the system
 \cite{SK02,CMB04,DB04,HDB05,ACCAD08,GBB08}. The dynamics of
 simple systems interacting with different environments
 has been experimentally studied using an all-optical
 device \cite{AlmeidaEtAl07,SallesEtAl08}.

 Some of these investigations \cite{SK02} have suggested that the entanglement
 of multi-qubit systems may become more robust as the number of qubits increases.
 It was found that, in some cases, the time it takes for the
 entanglement of a multi-qubit system to vanish due to decoherence
 increases with the number of qubits of the system.  Alas,
 recent results reported by Aolita \textit{et al.} \cite{ACCAD08} clearly indicate that,
 even in those cases where entanglement takes a long time to entirely
 disappear, it soon becomes too small to be of any practical use.
 Even worse, the larger the number of qubits of the system, the sooner
 this happens. Interestingly enough, the findings of Aolita \textit{et al.} are
 fully consistent with the fast decoherence processes invoked
 (within the decoherence program) to explain the
 quantum-to-classic transition \cite{S05}.

 Aolita \textit{et al.} \cite{ACCAD08} studied the decrease of entanglement
 of an $N$-qubits system, initially in a  $|GHZ\rangle$ state,
 that experiences decoherence through the interaction of each of its
 qubits with an independent environment. Following this interesting
 and promising line of enquiry, the aim of the present
 contribution is to explore the entanglement-robustness of
 highly (in some cases maximally) entangled multiqubit-states, i.e., the ones
 that may be of technological utility.

 The paper is organized as follows.  In Section II we briefly
 review the local decoherence models for multi-qubit systems
 that we use in the present work. In Section III we
 investigate the entanglement robustness of highly entangled
 multi-qubit states. In Section IV we consider the entanglement
 decay corresponding to initial states equivalent under
 local unitary transformations to the $|GHZ\rangle$ state.
 In Section V we investigate whether the decoherence
 process leads to multi-qubit states exhibiting bound
 entanglement. Finally, some conclusions are drawn in Section VI.

\section{Decoherence Models}
The systems under consideration in the present
study consist of an array on $N$ independent
qubits initially entangled due to a previous, arbitrary,
interacting process. Each qubit in the composite system is coupled
to its own environment; in this local environment formulation
there is no communication and the entanglement between the
subsystems cannot increase because of the locality of the involved
operations. We assume that all qubits are affected by the same
decoherence process. The dynamics of any of these qubits is governed
by a master equation from which one can obtain  a completely
positive trace-preserving map $\varepsilon$ which describes the
corresponding evolution: $\rho_i (t) = \varepsilon \rho_i (0)$.
In the Born-Markov
approximation these maps (or channels) can be described using its
Kraus representation

\begin{equation}
 \varepsilon_i \  \rho_i(0) = \sum_{j=1}^{M} E_{j \, i} \
 \rho_i (0) \  E_{j \, i}^\dag,
\end{equation}

\noindent where $E_j\,\,\,j=1,\ldots, M$ are the so called Kraus
operators, $M$ being the number of operators needed to completely
characterize the channel \cite{K83}. There are other approaches to
describe noisy channels such as the quantum Liouville equation
\cite{QJ08}. Using the Kraus operators formalism it is possible to
describe the evolution of the entire N-qubit system,

\begin{equation}
 \rho(t)= \varepsilon \ \rho(0) \  = \  \sum_{i...j} E_{i \, 1} \otimes \ldots
 \otimes E_{j \, N} \ \rho(0) \  [ E_{i \, 1} \otimes \ldots \otimes E_{j \, N}]^{\dag}.
\end{equation}

We consider the following five paradigmatic noisy channels

\subsubsection{Phase Damping}
This process describes the loss of quantum information with
probability $p$ without any exchange of energy. Physical examples
of this process are given by the random scattering of a photon while
traveling through a waveguide, or the perturbation of the
electronic states in an atom when interacting with distant
electrical charges \cite{NC00}. The Kraus operators for this
channel are

\begin{equation}\label{PD-kraus}
E_0 = \left(
        \begin{array}{cc}
          1 & 0 \\
          0 & \sqrt{1-p} \\
        \end{array}
      \right)
 ; \ \ \ \ E_1 \, = \left(
                       \begin{array}{cc}
                         0 & 0 \\
                         0 & \sqrt{p} \\
                       \end{array}
                     \right)
.
\end{equation}

\subsubsection{Depolarizing}

This one can be viewed as a process in which the initial state is
mixed with a source of white noise with probability $p$. Because the
channel is highly symmetric, all output states are unitarily
equivalent. For a $d$-dimensional quantum system, it can be
expressed as
\begin{equation}\label{D-map}
\varepsilon \, \rho \, = \, \frac{p}{d} \, I + (1-p) \, \rho
\end{equation}

\noindent where $I$ stands for the $d\times d$ identity matrix. The
Kraus operators for this process, including all Pauli matrices are

\begin{equation}\label{D-kraus}
E_0 \, = \, \sqrt{1 - p'} I \, ; \ \ \ E_i \, = \,
\sqrt{\frac{p'}{3}} \sigma_i
\end{equation}

\noindent where $p' = \frac{3p}{2}$, $i=1,2,3$, and $\sigma_i$ are
the corresponding Pauli matrices $\sigma_x, \sigma_y, \sigma_z$.

\subsubsection{Bit flip, phase flip, and bit-phase flip}

These channels represent all the possible errors in quantum
computation, the usual bit flip $0\leftrightarrow 1$, the phase
flip, and the combination of both, bit-phase flip. The corresponding
pair of Kraus operators $E_0-E_1$ for each channel is given by:

\begin{equation}\label{BF-kraus}
E_0 = \sqrt{1-p/2}I, E_1^i=\sqrt{p/2}\sigma_i;
\end{equation}
\noindent where $i=x$ give us the bit flip, $i=z$ the phase flip,
and $i=y$ the bit-phase flip. Following Salles {\it et.al}
\cite{SallesEtAl08}, the factor of 2 in Eq. (\ref{BF-kraus})
guarantees that at $p=1$ the ignorance about the occurrence of an
error is maximal, and as a consequence, the information about the
state is minimum.

\section{Robust maximally entangled states}
\subsection{Preliminaries}
In this section we study the decay of entanglement of maximally
entangled multiqubit pure states. There exist various measures that
aim to quantify and characterize different features of the
multipartite entanglement phenomenon. We  stress that by maximally
entangled states we mean those states
maximizing an appropriate entanglement measure under the requirement
that entanglement is (at least approximately) uniformly shared
among all the system's components. These states may seem easy to
characterize: all their reduced density matrices must be maximally
mixed. A good example of a state complying with this requirement is
the 3 qubit $|GHZ\rangle$ state. Alas, for systems with more than
three qubits the $|GHZ\rangle$ or {\it cat state} is no longer the
maximally entangled state (at least, not in the aforementioned
sense). The characterization and quantification of maximally
entangled states for systems of $N>3$ qubits has recently been the
focus of an intensive research activity
\cite{HS00,BH07,BSSB05,BPBZCP07,BCPP08,FFPP08}.

One of the most popular measures proposed to quantify such
multipartite entanglement is based on the use of a bipartite
measure, which is averaged over all possible bipartitions of the
system. It is mathematically expressed by:

\begin{eqnarray}\label{Ent-measure}
E &=& \frac{1}{[N/2]} \sum_{m=1}^{[N/2]} E^{(m)}, \\
E^{(m)} &=& \frac{1}{N_{bipart}^m} \sum_{i=1}^{N_{bipart}^m} E(i).
\label{Entsub}
\end{eqnarray}

\noindent Here, $E(i)$ stands for the entanglement associated with
one, single bipartition of the $N$-qubit system. The quantity
$E^{(m)}$ gives the average entanglement between subsets of $m$
qubits and the remaining $N-m$ qubits constituting the system. The
average is performed over the $N_{bipart}^{(m)}$ nonequivalent
bipartitions, which are given by

\begin{eqnarray}
N_{bipart}^{m} &=& \binom{N}{n}\textrm{     if }n \neq N/2,\\
N_{bipart}^{N/2} &=& \frac{1}{2} \binom{N}{N/2}\textrm{     if }n = N/2.
\end{eqnarray}

The total number of bipartitions is equal to

\begin{equation} \label{eq:ent.ncuts}
N_{cuts} = \sum_{i=1}^{[N/2]} N_{bipart}^{i} = 2^{N-1}-1.
\end{equation}

Different $E^{(m)}$ represent different entanglement features
of the state.  Two states may, for instance, share the
same value of $E^{(1)}$ and have very different values
of $E^{(2)}$. A state may even maximize $E^{(1)}$ while
exhibiting very low values of $E^{(2)}$. It is clear
that all the entanglement measures $E^{(m)}$ have to
be computed in order to describe the entanglement
properties of a multipartite state as comprehensively as
permitted by the present bi-partitions based approach.
However, it is also desirable to try to characterize
 with one single number the ``total'' amount of
entanglement associated with the state. Alas, any such
attempt will inevitably exhibit some degree of
arbitrariness for the very fact that multipartite
entanglement is a highly complex phenomenon whose
description cannot be ``compressed'' into
a single number. In the present work we adopted
the global multiqubit entanglement measure (\ref{Ent-measure}),
given by the average of the $[N/2]$ different $E^{(m)}$
associated with a state $\rho$. An alternative procedure
would be to compute the entanglement corresponding to
each of the $N_{cuts}$ possible bipartitions and then
evaluate directly their average, skipping the intermediate
step of computing the quantities $E^{(m)}$. However, this
scheme implies a strong bias towards the most balanced
bipartitions because they are more numerous than the
unbalanced ones (see table \ref{tb:eq.ncuts}).
We have performed numerical experiments using both the
aforementioned ways of computing the global amount
of entanglement of a multi-qubit state and found that
the main conclusions reported in the present work are
valid in both cases.

We will use the negativity as our bipartite measure
of entanglement because we are dealing with mixed states.

\vskip 0.2cm
\begin{table}[]
\begin{center}
\begin{tabular}{cc|c|c|c|c|c|}
\cline{3-7}
& & \multicolumn{5}{|c|}{N} \\ \cline{3-7}
& & 3 & 4 & 5 & 6 & 7 \\ \cline{1-7}
\multicolumn{1}{|c|}{\multirow{3}{*}{m}} &
\multicolumn{1}{|c|}{1} & 3 & 4 & 5 & 6 & 7    \\ \cline{2-7}
\multicolumn{1}{|c|}{} &
\multicolumn{1}{|c|}{2} & 0 & 3 & 10 & 15 & 21    \\ \cline{2-7}
\multicolumn{1}{|c|}{} &
\multicolumn{1}{|c|}{3} & 0 & 0 & 0 & 10 & 35    \\
\hline
\multicolumn{2}{|c|}{$N_{cuts}$} &  3 & 7 & 15 & 31 & 63 \\
\hline
\end{tabular}
\end{center}
\caption{Number of non-equivalent bipartitions as a function of the
number of qubits of the system $N$, and the dimension $m$ of the
smaller subsystem that results from the bipartition.}
\label{tb:eq.ncuts}
\end{table}

Any set of unitary local operations applied to a maximally entangled
state will result in a different state with the same amount of
entanglement. When a given decoherence model is applied to this
family of locally equivalent states quite different entanglement
evolutions may arise. Our interest is focused on maximally entangled
states that are able to retain during their dynamical evolution a
larger amount of entanglement $E$ than the one kept by other states
with the same initial  $E-$amount. Following the
robustness-definition introduced in Ref. \cite{BL05}, we will refer
to these states as {\it robust entangled ones} and use the notation
$|\Psi_{rob}^N \rangle$ for them, where $N$ is the number of qubits
of the system.
 We compute the entanglement dynamics of these states under the
influence of the various decoherence channels introduced in the
previous section, and compare such dynamics  with that of the
entanglement decay of the N-qubit $|GHZ\rangle$ and $|W\rangle$
states.

To find initial states $\rho(0)$ exhibiting robust entanglement
under a decoherence channel $\varepsilon$ we use an iterative
numerical search scheme akin to the well-known simulating annealing
algorithm. During the search process a series of initial states
$\rho^{(i)}(0),\,\,\,i=1,2,3,\ldots$ of increasing robustness is
generated (here the word ``initial" refers to the
decoherence-evolution, not to the search algorithm). At each step of
the search process, a (random) new state $\rho^{(k+1)}(0)$ slightly
different from the previous one $\rho^{(k)}(0)$ is generated. If the
entanglement of the state $\varepsilon \rho^{(k+1)}(0)$ is larger
than the entanglement of $\varepsilon \rho^{(k)}(0)$, the new state
$\rho^{(k+1)}(0)$ is kept. Otherwise, the new state is rejected and
a new, tentative state is generated. This iterative process is
repeated until it converges to an initial state  $\rho(0)$ maximizing
the entanglement of the corresponding evolved state $\varepsilon
\rho(0)$. Notice that the entanglement that we are maximizing is not
the entanglement of  $\rho(0)$ itself, but the entanglement of the
state $\varepsilon \rho(0)$ resulting from the action of the
decoherence channel $\varepsilon$ upon $\rho(0)$. Since we use the
negativity as our measure of entanglement, the quantity to be
maximized is the negativity of the mixed state obtained after
applying the decoherence channel $\varepsilon$ to the initial state
$\rho(0)$.
Instead of studying states $\rho(0)$ that maximize the
entanglement $E[\rho(t)]$ present at a later time $t$,
 one could consider initial states $\rho(0)$ optimizing the ratio
$E[\rho(t)]/E[\rho(0)]$. However, this ratio may adopt large
values (that is, values close to one) for states with little
initial (and final) entanglement that are not interesting
as resources for quantum information processes.

\subsection{Present results}
As a result of the above numerical search process we were able to
ascertain that some of the maximally entangled multiqubit states
that can be found in the literature are already robust. However, for
some systems we have found new states exhibiting a higher degree of
entanglement robustness than the one associated with previously
known states.

\begin{figure}
\begin{center}
\vspace{0.5cm}
\includegraphics[scale=0.75,angle=0]{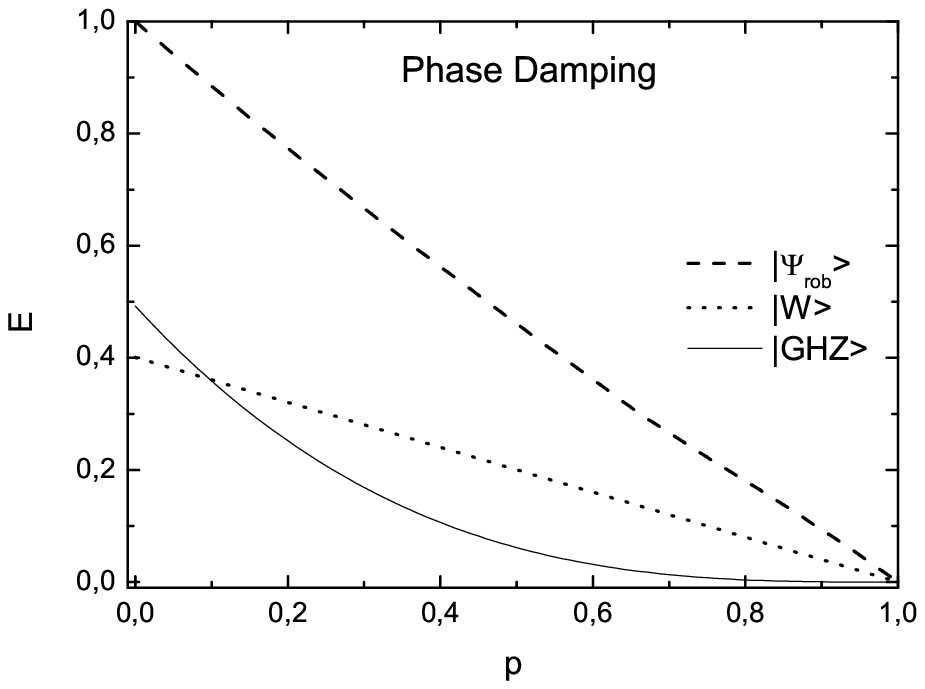}
\vspace{0.5cm}
\includegraphics[scale=0.75,angle=0]{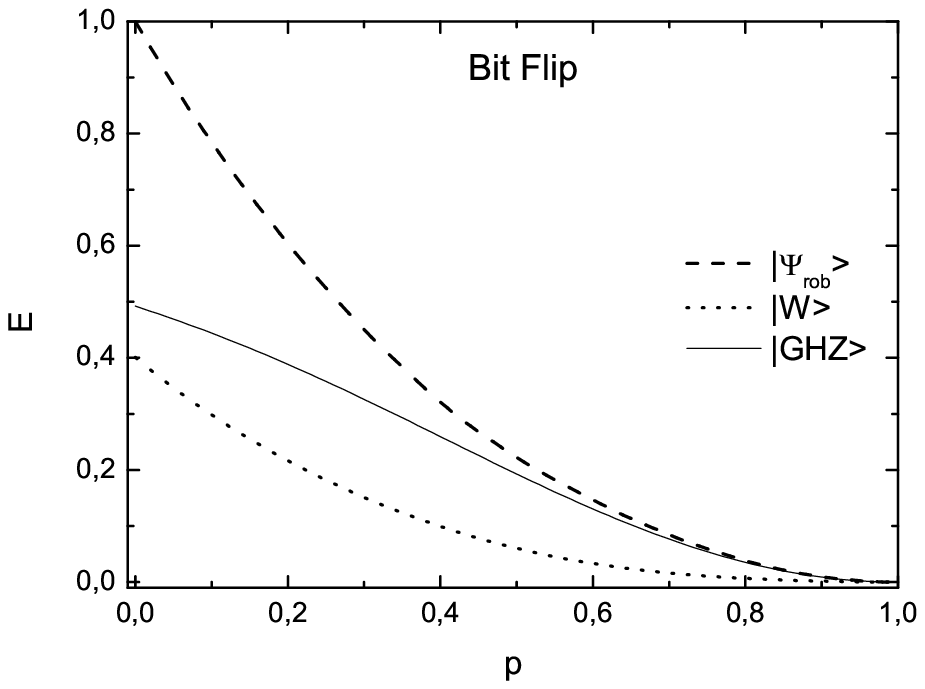}
\caption{Entanglement evolution of the 6 qubits representative
states under phase damping and bit flip decoherence
models. \label{fig_states}}
\end{center}
\end{figure}

At this point we begin the explanation of our novel results for the
highest dimensional system studied in this work, that composed of 6
qubits. We find that the robust state $|\Psi_{rob}^6 \rangle$ turns
out to be precisely that maximally entangled state encountered  by
some of the authors on a previous work \cite{BPBZCP07}. No  pure
state of 6 qubits has been found that evolves to a mixed state with
a higher amount of entanglement. In Fig. \ref{fig_states}a one
observes that for the PD channel the negativity of $|\Psi_{rob}^6
\rangle$ is almost a straight line, and a similar behavior is
observed for the $|W\rangle$ state. Instead,   $|GHZ\rangle$ quickly
loses entanglement vis-a-vis the preceding examples. A qualitatively
similar decay of entanglement is obtained in the case of  the PF
channel. For the BF-one (plotted in Fig. \ref{fig_states}b) and also
for the BPF channel, the entanglement of the $|GHZ\rangle$ is more
robust than that for the preceding channels. It remains always
higher than the one associated to $|W\rangle$, although lower than
that of $|\Psi_{rob}^6 \rangle$. Finally, the depolarizing channel
is the only one for which the initial entanglement seems to be the
only important factor in our scenario, the entanglement-decay of all
our states (equally entangled initially) sharing such a behavior.
This means that any state belonging to the set of initially maximally
entangled states, those equivalent under LU operations to
$|\Psi_{rob}^6 \rangle$, is robust according to the depolarizing channel.

For lower dimensional systems the overall picture resembles that of
the 6 qubits case, but some particularities must be pointed out. For
5 qubit systems, the robust state $|\Psi_{rob}^5 \rangle$ that we
find is not so good as the one for 6 qubits. Its entanglement decay
is just maximal for four out of the six channels under analysis. For
the two channels (BF and BPF) for which is not maximal  its
entanglement becomes lower than that of other states for large $p$
values. The explicit form of $|\Psi_{rob}^5 \rangle$ is given by

\begin{eqnarray}\label{rob5-state}
| \Psi_{rob}^{5} \rangle \, = \, \frac{1}{\sqrt{16}} \, [
-|5\rangle + |6\rangle - |9\rangle + |10\rangle + |17\rangle +
|18\rangle -|29\rangle - |30\rangle  + \nonumber \\ \imath (-
|0\rangle + |3\rangle - |12\rangle + |15\rangle - |20\rangle -
|23\rangle + |24\rangle + |27\rangle ) ].
\end{eqnarray}
In this equation the ket $|i\rangle$ denotes a member of the
computational basis for a system of 5 qubits, which can be
identified by expressing the integer $i$ in binary notation. For
instance, $|3\rangle$ is a short hand notation for $|00011\rangle$.

For systems of 4 qubits, it was proved in \cite{HS00} that a pure
state exhibiting the theoretically maximum amount of entanglement
(that is, having all its marginal density matrices maximally mixed)
does not exist. In Ref. \cite{HS00},  a promising candidate for the
maximally entangled status was also proposed, namely,

\begin{equation}\label{HS-state}
| \Psi_{rob}^{4} \rangle \, = \,|HS\rangle \, = \,
\frac{1}{\sqrt{6}} \Bigl[ |1100\rangle+ |0011\rangle + \omega
\Bigl(|1001\rangle + |0110\rangle \Bigr) + \omega^2
\Bigl(|1010\rangle + |0101\rangle\Bigr) \Bigr],
\end{equation}
\noindent with $\omega= -\frac{1}{2}+ \imath \frac{\sqrt{3}}{2}$.
This conjecture has later received support from several numerical
studies \cite{BPBZCP07,BCPP08,FFPP08}. The entanglement decay of
(\ref{HS-state}) resembles that of $|\Psi_{rob}^6 \rangle$ and its
entanglement is always larger than that of any other state tested in
our samplings.

Contrary to what could be expected, the 3 qubit case is the
most complex one. The small number of parameters needed to characterize a
3 qubit state is low enough to perform the numerical optimization process
in a very short time, allowing us to identify states which are robust for
different channels. The main difference between 3-qubit systems
and the higher dimensional systems considered previously
is that there is no 3-qubit robust state that simultaneously
maximizes the entanglement for all the decoherence channels.
All 3-qubit states optimizing the entanglement-robustness for a
given channel quickly lose their entanglement when evolving under
other decoherence maps. The $|GHZ\rangle$
provides a clear illustration of the peculiar
features exhibited by three qubit systems. The entanglement of this
state is robust under the action of the BF channel. However, for
example, in the BPF channel's instance, the entanglement of the
$|GHZ\rangle$ remains large for small $p$ values, but it then decays
below the expected pattern until it almost vanishes for $p \simeq
0.7$. A similar behavior occurs with other channels, even when using
alternative states which are robust for some channels but whose
entanglement very quickly decays under the action of other
decoherence processes.

\section{Decoherence of initial States Equivalent Under
Local Unitary Operations to the $|GHZ\rangle$ State}

 In this Section we are going to consider the decoherence
 behavior of systems initially prepared in states equivalent
 (under local unitary operations) to the $|GHZ\rangle$ state.
 In particular, we are going to determine (for different
 decoherence channels) which one of the aforementioned states
 exhibits the most robust entanglement.

 By recourse to a numerical survey of the behavior
 of states equivalent to the $|GHZ\rangle$ one,
 we found that (among the alluded states) the
 one exhibiting the most robust entanglement
 for the phase damping and phase flip
 decoherence channels is the $N$-qubits state

 \begin{eqnarray} \label{elsime}
 | H\rangle_N \, &=& \, U_{H} \otimes \ldots \otimes U_{H} |GHZ\rangle \cr
 \, &=& \, \frac{1}{\sqrt{2}}\left(|+\rangle^{\otimes N} -
 |-\rangle^{\otimes N}\right),
 \end{eqnarray}

 \noindent
 obtained by applying the Hadamard gate  $U_H$ on each of the qubits
 of an $N$-qubits $|GHZ\rangle$ state. In equation (\ref{elsime}),
 $|+\rangle$ and $|-\rangle$ stand, respectively, for
 states $\frac{1}{\sqrt{2}}(|0\rangle+|1\rangle)$
 and $\frac{1}{\sqrt{2}}(|0\rangle -|1\rangle)$.

The decay of the amount of entanglement
 corresponding to various initial four-qubits states
  is depicted in Fig. 2 for the
 phase damping, depolarizing, bit flip, and bit phase
 flip decoherence channels, respectively. The four-qubits states
 considered in Fig. 2 are the $|GHZ\rangle$,
 the $|H\rangle_4$ provided by eq. (\ref{elsime}),
as well as 1000 states generated applying
 random local unitary transformations to the
 $|GHZ\rangle$ states. We also plot the decay of entanglement
for the robust state $|HS\rangle$ previously introduced
in eq. (\ref{HS-state}). Excepting the $|HS\rangle$
 state, all these states are equivalent under
 local unitary transformations to $|GHZ\rangle$.
As we mentioned in the previous section, the $|HS\rangle$
state is the most robust 4-qubit state for all the considered channels.
 We note that for the depolarizing channel the decay of entanglement
only depends on the initial amount of entanglement.
The decay of entanglement for the $|GHZ\rangle$ and $|H\rangle_4$
states under the PF, BF, and BPF channels is somehow equivalent,
in the sense that the decay for the $|GHZ\rangle$ for the
PF is the same than the decay of the $|H\rangle_4$ for the BF.
This is the reason why we only plot one of these channels,
as the other one is equivalent. This behavior is easily
explained because the unitary transformation $U_H$ that
maps the state $|GHZ\rangle$ into the $|H\rangle_4$
is the same that transforms the BF channel into the PF.

%
%

\begin{figure}
\begin{center}
\vspace{0.5cm}
\includegraphics[scale=0.65,angle=0]{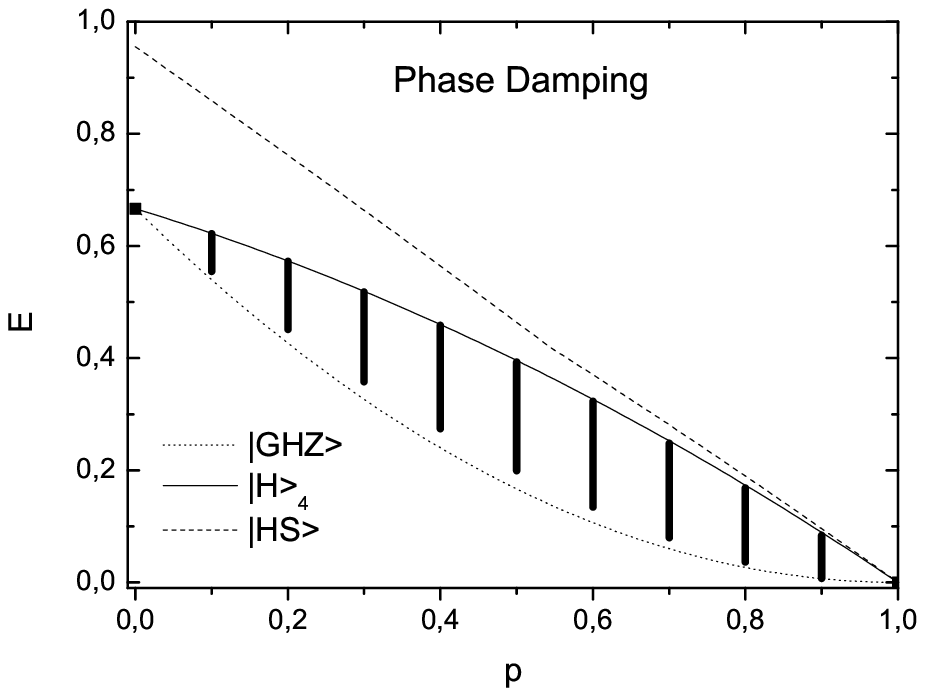}
\vspace{0.5cm}
\includegraphics[scale=0.65,angle=0]{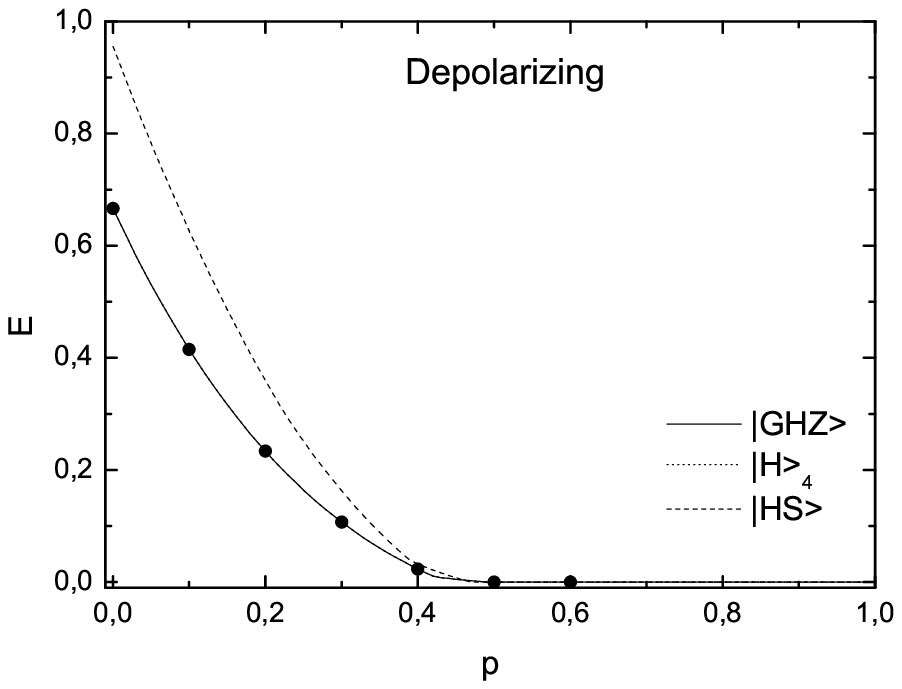}
\vspace{0.5cm}
\includegraphics[scale=0.65,angle=0]{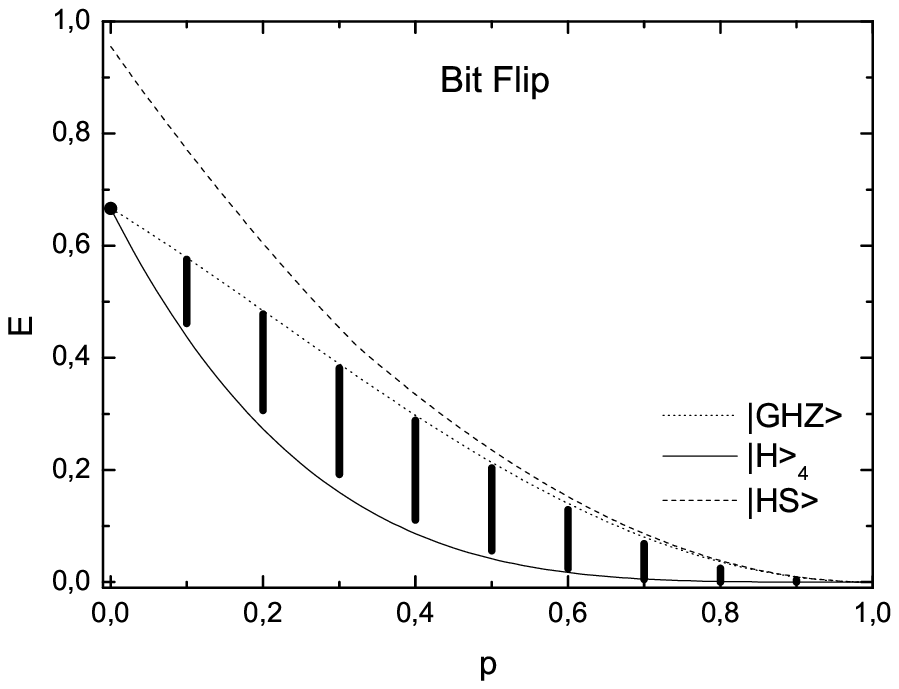}
\vspace{0.5cm}
\includegraphics[scale=0.65,angle=0]{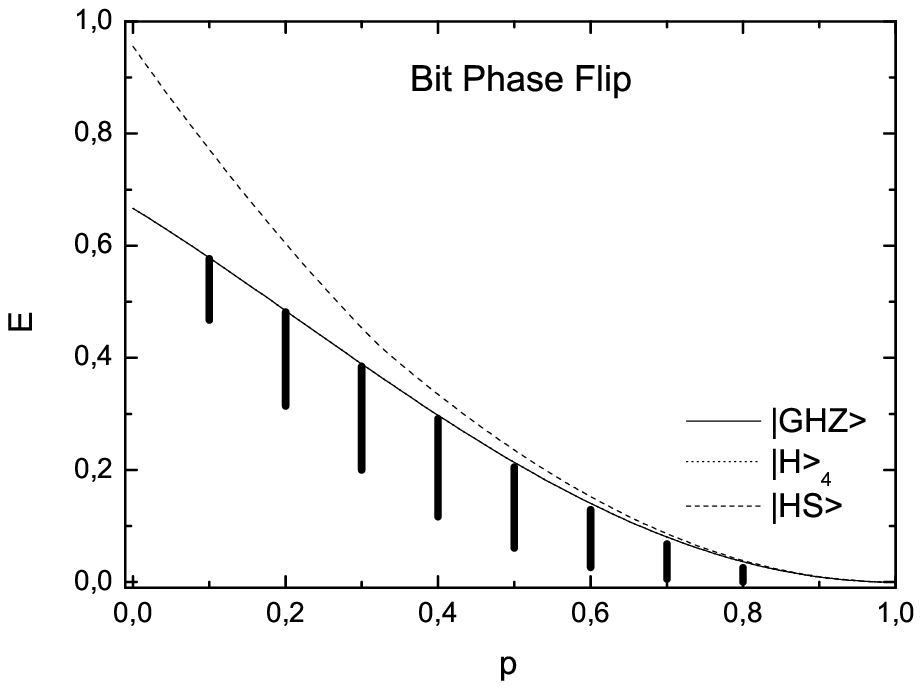}
\caption{$|H\rangle_4$ vs. $|GHZ\rangle$ states for a 4 qubits system. In all
subfigures the vertical lines correspond to the entanglement dynamics for 1000
states obtained from the $|GHZ\rangle$ by applying unitary local
transformations.\label{fig_sym}}
\end{center}
\end{figure}

 It is worth noticing that the states
 $|H\rangle_N$ can be defined in terms
 of the recurrence relation

 \begin{eqnarray} \label{recurrencia}
 |H\rangle_N\ \, &=& \,
 \frac{1}{\sqrt{2}} \Bigl( |0\rangle \otimes |H\rangle_{N-1} +
 |1\rangle \otimes |{\bar H}\rangle_{N-1} \Bigr), \cr
 |{\bar H}\rangle_N \, &=& \,
 \frac{1}{\sqrt{2}} \Bigl( |1\rangle \otimes |H\rangle_{N-1} +
 |0\rangle \otimes |{\bar H}\rangle_{N-1} \Bigr),
 \end{eqnarray}

 \noindent
 with

 \begin{eqnarray}
 |H\rangle_2 \, &=& \, \frac{1}{\sqrt{2}} (|00\rangle + |11\rangle), \cr
 |{\bar H}\rangle_2 \, &=& \, \frac{1}{\sqrt{2}} (|01\rangle + |10\rangle)
 \end{eqnarray}

\section{Bound entanglement}

\begin{figure}
\begin{center}
\vspace{0.5cm}
\includegraphics[scale=0.75,angle=0]{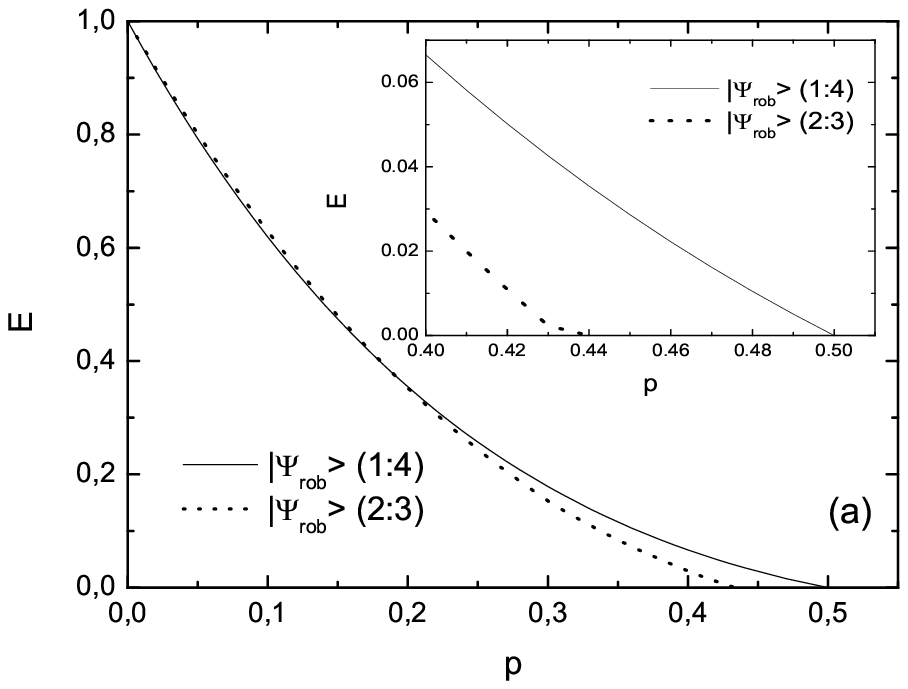}
\includegraphics[scale=0.75,angle=0]{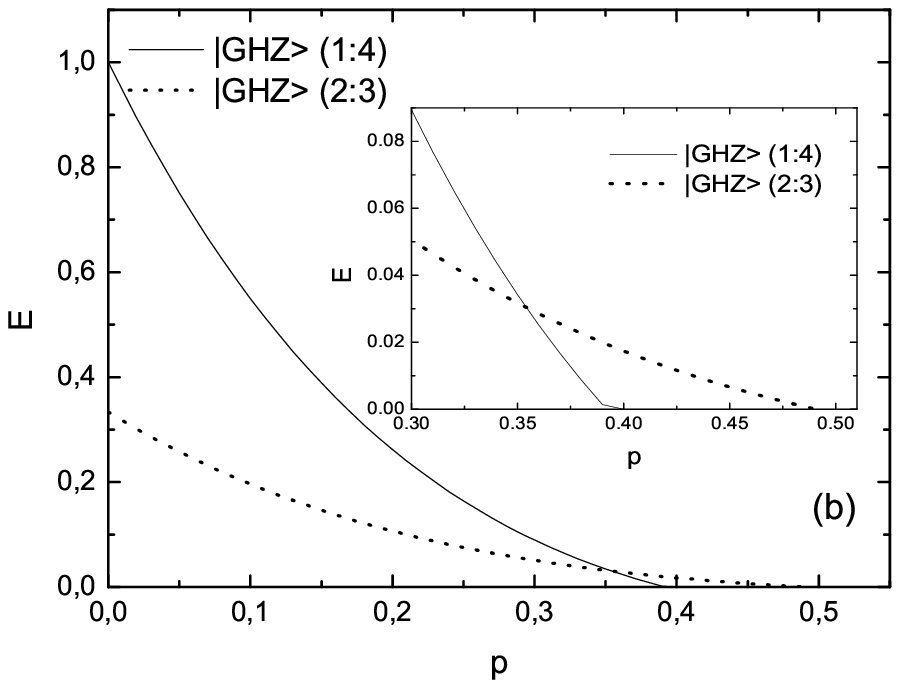}
\caption{Entanglement evolution of different bipartitions for the
robust (a) and GHZ (b) state of 5 qubits for the depolarizing
channel. The insets show a magnification of the region in which
the entanglement vanishes. \label{fig-BE}}
\end{center}
\end{figure}

A multipartite entangled state has bound entanglement if it 
is globally entangled but is separable according to any of 
its possible bipartitions \cite{DC00}. Aolita {\it et al.} 
investigated the existence of bound entanglement
in the final states resulting from  decoherence processes acting
upon initial states of the generalized $|GHZ\rangle$ form
\cite{ACCAD08}. They found that the action of the depolarizing
channel on a generalized $|GHZ\rangle$ state leads, for a certain
range of $p$-values, to states for which the entanglement of the
most unbalanced partitions $1:N-1$ vanishes while the state is still
entangled according to the most balanced partitions, as for instance
$N/2:N/2$. This kind of states are said to have ``bound
entanglement" because no entanglement can be distilled from them by
means of local operations, notwithstanding the fact that the system
is globally entangled. 

We are now going to investigate the presence of bound entanglement
in the $N$-qubit states generated by the decoherence process when the
initial states are the robust ones introduced in the previous Sections.
 In order to address this issue  we calculate (on the states
resulting from the action of the decoherence channel) the
negativity corresponding to different partitions of the system. We
concentrate our efforts on the depolarizing environment, because
in this case the phenomenon of entanglement's sudden death occurs 
in a finite time. Fig. \ref{fig-BE} shows that bound entanglement 
is observed both for the robust $|\Psi_{rob}\rangle$ and $|GHZ\rangle$ states. 
Actually, for all the initial states studied in this work (including a 
sample of 1000 randomly generated states) the decoherence process
leads to states endowed with bound entanglement. This typical 
behaviour indicates a remarkable capacity of the environment 
to create bound entanglement. However, special initial states 
exhibit interesting differences when the decay of entanglement associated with 
different types of bipartitions is considered. For the robust states $|\Psi_{rob}\rangle$ 
introduced in this work, as well as for all the random states that we have 
generated, the entanglement of the most balanced partitions $N/2:N/2$ vanishes 
sooner than the entanglement corresponding to the most unbalanced partitions $1:N-1$. 
This is in contraposition to what happens with the generalized 
$|GHZ\rangle$ states, as was found by Aolita {\it et al}. The 
$|W\rangle$ and the $|H\rangle_{N}$ states also share this 
uncommon behaviour: the $p$-values for which the 
entanglement of the most unbalanced partitions $1:N-1$ 
vanish corresponds to states that are still entangled 
according to more balanced partitions.

\section{Conclusions}

In the present effort we investigated the decay of the amount of
entanglement of a multi-qubit system experiencing a decoherence
process. We considered models of decoherence characterized by
independent environments respectively interacting with each of the
system's qubits. We performed a systematic numerical search of the
initial pure states exhibiting the most robust entanglement under
decoherence scenarios described by five different channels: phase
damping, depolarizing, bit flip, phase flip, and bit-phase flip. For
systems of 4 qubits we found that the state having the most robust
entanglement is the $|HS\rangle$ state introduced by Higuchi and
Sudbery \cite{HS00}, which was conjectured to correspond to a global
maximum of entanglement for four-qubits systems
\cite{HS00,BSSB05,BH07,BPBZCP07,BCPP08}. In the case of six qubits
the state maximizing entanglement robustness again coincides with an
already known state of maximum entanglement that had been the
subject of previous studies \cite{BPBZCP07,BCPP08}. The behavior of
the states optimizing robustness for 4 and 6 qubits is similar: both
states maximize entanglement-robustness for each of the six
decoherence channels considered in the present work.

We also were able to determine a five-qubits state with highly
robust entanglement. However, this state exhibits optimal robustness
only for four of the decoherence channels considered here. Under the
effect of the remaining decoherence channels, and for large enough
values of the ``time" parameter $p$, {\it this initial state}
evolves into a state with less entanglement than that  obtained from
{\it other} initial states. When this happens, however, the
entanglement exhibited by the system (regardless of the initial
state) is already too small to be of any practical use.

Unexpectedly, the case of three-qubits turned out to be the most
complex one. Our numerical results indicate that for three-qubits
systems there is no state simultaneously maximizing the
entanglement-robustness for all (or, at least, for
most of) the decoherence channels studied here. In other words, each
state exhibiting optimal robustness for one of the channels was
found to be non optimal for some of the other channels.

 We also discussed a restricted version of the robustness'
 optimization problem involving only states equivalent to the $|GHZ\rangle$
 one under
 local unitary transformations.
 A state $|H\rangle_N$ obtained by applying the Hadamard gate
 to each of the qubits of a multi-qubit system
 in the $|GHZ\rangle$ state was found to play
 an interesting role. This special state maximizes the
 entanglement-robustness for four decoherence channels.

 Finally, we studied the phenomenon of bound entanglement 
determining whether initially entangled multi-qubit states evolve (as the
 decoherence process takes place) into states
 characterized by bound entanglement. We focused
 our attention on the depolarizing channel,
 this being the channel leading to entanglement
 sudden death in a finite time. We found that the ability of the 
environment to create bound entanglement is a quite general feature that 
appears for all the states studied in this work.

\begin{acknowledgements}

This work was partially supported by the MEC grant FIS2005-02796
(Spain), by the Project FQM-2445 of the Junta de Andalucia (Spain),
by FEDER (EU), and by Conicet (Argentine Agency). AB acknowledges
support from MEC through FPU fellowship AP-2004-2962 and APM
acknowledges support of MEC contract SB-2006-0165.
\end{acknowledgements}

\end{document}